\documentclass[pre,reprint, showpacs]{revtex4-1}
\usepackage{amsmath, graphicx}
\begin{document}
\title{Spinodal fractionation in a polydisperse square well fluid}

\date{\today}

\author{J.~J.~Williamson}
\email{pyjjw@leeds.ac.uk}
\author{R.~M.~L.~Evans}
\affiliation{Soft Matter Group, School of Physics and Astronomy, University of Leeds, Leeds LS2 9JT United Kingdom}

\begin{abstract} 
Using Kinetic Monte Carlo simulation, we model gas-liquid spinodal decomposition in a size-polydisperse square well fluid, representing a `near-monodisperse' colloidal dispersion. We find that fractionation (demixing) of particle sizes between the phases begins asserting itself shortly after the onset of phase ordering. Strikingly, the direction of size fractionation can be reversed by a seemingly trivial choice between two inter-particle potentials which, in the monodisperse case, are identical -- we rationalise this in terms of a perturbative, equilibrium theory of polydispersity. Furthermore, our quantitative results show that Kinetic Monte Carlo simulation can provide detailed insight into the role of fractionation in real colloidal systems.
\end{abstract}


\maketitle

\section{\label{sec:intro}Introduction}

Colloidal dispersions consist of particles of mesoscopic size dispersed in a fluid solvent, a  definition that embraces a huge variety of natural and synthetic substances. Milk, blood, viruses, paints, cosmetics and even radioactive waste are all colloidal, in that they are made up of some continuous background solvent in which one or many species of colloidal particle are dispersed. Since the microscopic dynamics of the colloidal particles are dominated by thermal fluctuations in the solvent, they can be studied using familiar thermodynamic and statistical mechanical approaches \cite{Onsager1949}. Therefore, as well as being studied for their own sake, colloids are seen as a useful analogue of ordinary molecular fluids and -- especially due to the relatively accessible length and time scales involved \cite{Prasad2007} -- as an experimental testing ground for more general theories  \cite{Suresh2006, Wette2009}. Nevertheless, the analogy between molecular and colloidal fluids is not exact. One key difference is that colloidal particles are inevitably \textit{polydisperse}, i.e. they exhibit continuous variation in particle properties such as size, charge or shape. Polydispersity leads to complex phase behaviour and complicates the theoretical and experimental treatment of complex fluids \cite{Allard2004, Sollich2011, Fasolo2004, Liddle2011, Sollich1998, Warren1998}. Increasing the breadth, precision and utility of our understanding of soft matter physics thus relies in part on a deeper understanding of the behaviour of polydisperse systems.
	
Much existing work has concerned itself with the equilibrium thermodynamic effects of size-polydispersity \cite{Evans1998, Liddle2011, Evans2000, Sollich2009, Sollich2011, Fasolo2004, Wilding2004, Fantoni2006, Bartlett1999}. Polydispersity is predicted and/or observed to cause destabilisation of the solid phase,  fractionation (demixing) of particle sizes between phases \cite{Wilding2004, Evans1998} and an apparent `terminal polydispersity' beyond which the system must split into multiple solid phases to reach equilibrium \cite{Sollich2011}. 

In contrast, theoretical work on the \textit{dynamics} of polydisperse phase behaviour  \cite{Warren1999, Evans2001} is rare, and the dynamical picture of how polydisperse systems actually do or do not arrive at equilibrium is far from complete \cite{Zaccarelli2009, Williams2008, Martin2003, Martin2005, Schope2007}. Moreover, in such studies as do exist, polydispersity is often introduced only to inhibit crystallisastion so that high density amorphous states can be studied \cite{Auer2001, Weysser2010}. 

Fractionation is thought to be involved in the nucleation of crystals from polydisperse fluids \cite{Martin2003, Martin2005, Schope2007}, with frustration of this mechanism held responsible for the experimental inaccessibility of some complex equilibrium phase diagrams \cite{Liddle2011}. In practical applications, fractionation of particle species between separating phases may be intended, for example to purify a mixture, or unintended. In either case, it is important to consider consequent effects on related properties (composition, crystal lattice parameter etc.) in the resultant phases. 

In the present work we address the issue by dynamically simulating gas-liquid spinodal decomposition in a size-polydisperse colloidal fluid. We observe fractionation phenomena on simulation timescales, and uncover a surprising example of the sensitivity of such phenomena to the details of particle interactions, which we explain by appealing to equilibrium theory.

\section{\label{sec:simulation}Simulation Setup}

The system studied in this work consists of either $N = 30000$ or $N = 4000$ spherical particles whose diameters $d_{n}$ are drawn from a Bates (pseudo-Gaussian) distribution of polydispersity (defined as ratio of the standard deviation to the mean) $\sigma = 0.06$ or $\sigma = 0.2$ \footnote{The Bates distribution is a sum of \protect{$n$} random numbers uniformly distributed on the unit interval -- in our implementation \protect{$n = 4$}.}. The simulation cell is cubic with periodic boundaries. We set the length unit as the mean hard core diameter $\langle d \rangle$. In order to mimic the qualitative features of any attractive interaction in the colloid, the hard particle cores are surrounded by a pairwise additive square well potential of (mean) range $\lambda = 1.15$ and depth $u = 1.82$, where the energy unit is $k_{B}T$. In the monodisperse limit, these parameters and the overall `parent' volume fraction $\phi_{p} = 0.229$ place the system in the middle of the gas-liquid coexistence region \cite{Liu2005}, where phase separation would be expected to proceed by spinodal decomposition. For the value of $\lambda$ used here, the resulting gas-liquid coexistence is metastable with respect to fluid-solid phase separation \cite{Pagan2005} which, on the timescale simulated, we do not observe.

Note that there is a choice to be made in precisely how the specific attraction range between any two particles $i$ and $j$ should depend upon their size. We have simulated two possibilities. The `scalable' dependence is motivated by its simplifying effect on the relevant theory viz. the equilibrium effects of fractionation \cite{Evans2000}:

{\begin{equation}
V_\textrm{{scal}}(r) = 
\begin{cases}
\infty & \text{if } r\leq d_{ij} \\
-u & \text{if } d_{ij} < r\leq \lambda d_{ij}\\
0 & \text{if } r > \lambda d_{ij}
\end{cases}
\label{eqn:scalablesquarewell}
\end{equation}}

\noindent where $d_{ij} = (d_{i} + d_{j})/2$. The `non-scalable' dependence is chosen for its similarity to depletion attractions \cite{Lekk1992}, in which the attraction range depends on the hard core diameter via an additive constant, e.g. the radius of gyration of a dissolved polymer:

{\begin{equation}
V_\textrm{{non-scal}}(r) = 
\begin{cases}
\infty & \text{if } r\leq d_{ij} \\
-u & \text{if } d_{ij} < r\leq d_{ij} + (\lambda - 1)\\
0 & \text{if } r > d_{ij} + (\lambda - 1)\\ 
\end{cases}~.
\label{eqn:nonscalablesquarewell}
\end{equation}}

\noindent In the monodisperse case, $d_{ij} = 1$ for all particle pairs so that the two definitions are strictly identical. Even in the polydisperse case the \textit{mean} attraction range is unaffected by the choice of dependence, but it will nevertheless turn out that this easily-overlooked detail can make a radical difference to the fractionation of the system.

Kinetic (or `dynamic') Monte Carlo (MC) simulation is distinct from equilibrium MC methods in that it is restricted to dynamically realistic trial moves, i.e. small stochastic `hops' representing the movement of particles performing Brownian motion. This is in contrast to equilibrium algorithms in which unphysical moves (e.g. cluster rearrangements or biased sampling) may be used to speed up equilibration so long as detailed balance is respected. For small step sizes, dynamical results from Kinetic MC simulation compare well with those of Brownian dynamics \cite{Sanz2010}. The simplicity of Kinetic MC and its natural suitability for representing inertia-free stochastic motion has resulted in a number of applications in soft matter simulation \cite{Auer2001, Sanz2007, Auer2005}.

In our implementation, trial particle moves are drawn from a Bates distribution centred on zero, and the mean step size for particle $i$ is $\Delta\sqrt{\langle d \rangle/d_{i}}$, so as to give the correct dependence of the short time Stokes-Einstein diffusion coefficient upon particle diameter. (The step parameter is set to $\Delta = 0.02$, and reducing this value does not alter the results presented here). The time unit is the time $t_{d}$ taken for a free particle of diameter $\langle d \rangle$ to diffuse a distance equal to its own diameter.

The Metropolis acceptance criterion, popular for equilibrium Monte Carlo (MC) simulations, assigns an equal probability (unity) to all moves which do not increase the system's energy, even those which \textit{decrease} it. In order to more naturally represent the dynamics of attractive particles, we instead employ an acceptance probability \cite{Novotny2001, Martin1977}
	
{\begin{equation}
P_{\textrm{accept}} = \frac{\exp({-E_{\textrm{new}}})}{\exp({-E_{\textrm{new}}}) + \exp({-E_{\textrm{old}}})}
\label{eqn:glauber}
\end{equation}}

\noindent to move from a configuration with energy $E_{\textrm{old}}$ to one with $E_{\textrm{new}}$. Moves that conserve energy are now accepted with probability 0.5, so that moves that bring a particle into the reach of attractive wells can be assigned a higher probability, representing more rapid movement down a potential gradient. 

The Kinetic MC algorithm allows large systems to be simulated on a physically relevant timescale, necessary for measurement of the required statistics for fractionation. However, as in comparable Molecular or Brownian dynamics algorithms, hydrodynamic interactions (HI) are neglected. Given the importance of HI in spinodal decomposition, this requires some comment. In an approximate theoretical scheme, for example, HI have been shown \cite{vandeBovenkamp1994} to enhance longer-wavelength decomposition while suppressing shorter wavelengths. Measuring the effect of such corrections on fractionation in a polydisperse system would require for instance explicit treatment of the solvent via Molecular dynamics, or a combined Brownian dynamics/lattice Boltzmann method, considerably increasing computational expense. 

Nevertheless, we expect that the phase separation observed here is at least qualitatively representative of spinodal decomposition in a real system and, particularly given that HI can be screened (e.g. by direct interactions in dense systems \cite{Riese2000}), constitutes a relevant and important baseline in the low-HI limit. Given this, and since our objective is to collect good statistics for the observation of fractionation, we leave the incorporation of HI to future work.

\begin{figure}[floatfix]
\includegraphics[width=7.2cm]{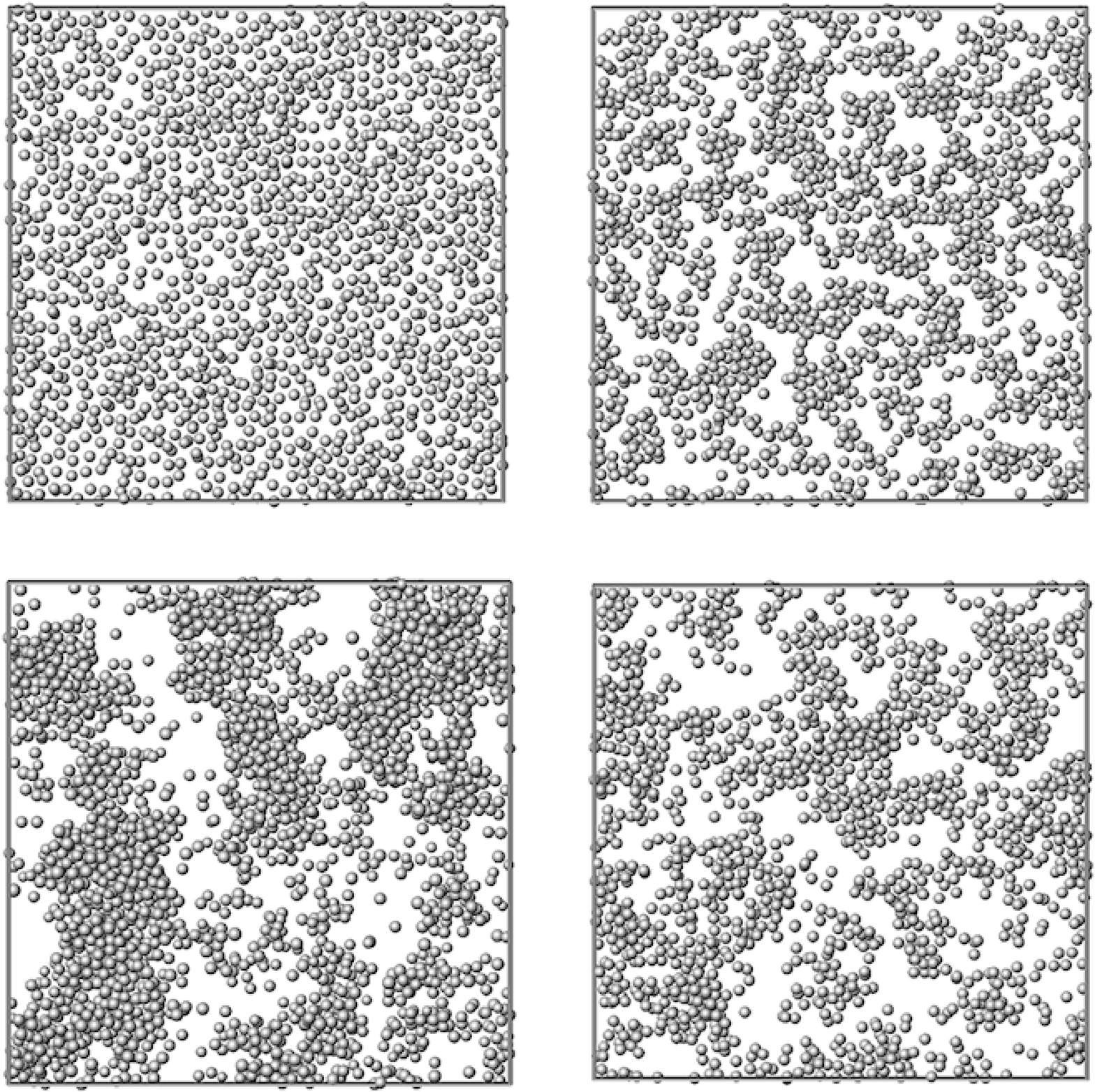}
\caption{\label{snapshots} Illustrative simulation snapshots at (clockwise from top-left): $t = 0$, $t = 4$, $t = 64$ and $t = 760$ for a slice 3 units wide through the system, showing the coarsening spinodal texture.}
\end{figure}

\begin{figure}[floatfix]
\includegraphics[width=7.2cm]{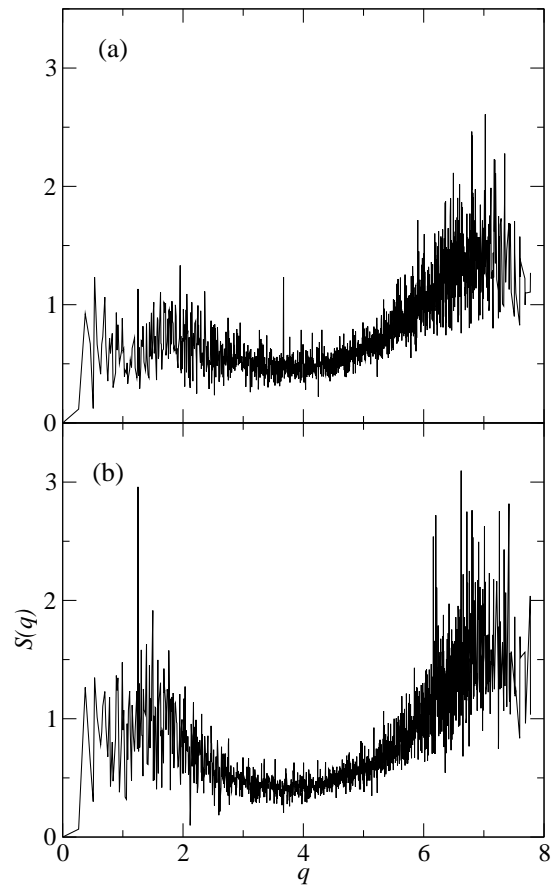}
\caption{\label{sk_long} Structure factor $S(q)$ for one simulation trajectory, averaged over a) $t = 1 - 2$ and b) $t = 3 - 4$. The usual fluid peak is present at $q \approx 7$, while a peak corresponding to the gas-liquid texture appears at $q \approx 2$ and shifts to a lower wavenumber as it grows, indicating coarsening of the two phases.}
\end{figure}

\begin{figure}[floatfix]
\includegraphics[width=7.2cm]{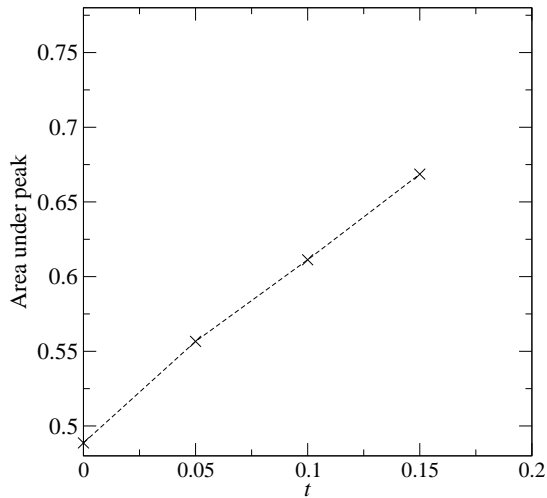}
\caption{\label{peak_early} The area under the low-$q$ peak, found by integrating from $q = 0.26$ to $q = 2.56$, averaged over 3 independent quenches, for times immediately after the quench. Although background noise, indicated by the value at $t=0$, is strong, the area under the peak is clearly increasing from these very early times, consistent with spinodal decomposition. The dashed line is a guide to the eye.}
\end{figure}

\section{\label{sec:spinodal decomposition}Spinodal decomposition}

The system is initialised as an amorphous fluid, whereafter the square well attraction is turned on (defining an instantaneous quench at $t = 0$) and the system evolved according to the Kinetic Monte Carlo algorithm outlined above. In addition to direct observation through time (FIG.~\ref{snapshots}), the development of a low-$q$ peak in the structure factor $S(q)$ is used to track the gas-liquid phase separation (FIG.~\ref{sk_long}). As shown in FIG.~\ref{sk_long}, this peak has already started to shift to lower $q$ after 2 or 3 $t_d$, indicating coarsening of the gas-liquid texture. To confirm the spinodal nature of the separation, FIG.~\ref{peak_early} shows the integrated area under the peak increasing on a timescale of hundredths of $t_d$. We detect no `lag' period associated with nucleation-driven phase separation, so we conclude that the phase separation is indeed spinodal in nature, as expected for our parameters \cite{Liu2005}.

Other dynamical observations, to be presented elsewhere, are in agreement with Ref.~\cite{Fasolo2005} in suggesting that, as in the monodisperse case, the gas-liquid coexistence resulting from the spinodal decomposition is in turn metastable with respect to the growth of a crystalline phase that would coexist with a tenuous vapour. On the timescales observed here and in the absence of a template to seed nucleation, we do not observe crystals.

\section{\label{sec:fractionation}Fractionation}

To check for fractionation, a per-particle distinction between the gas and liquid phases is required. We count a particle as liquid if it has more than 12 neighbours within 2 length units, this threshold corresponding to a local number density in the middle of the gas-liquid binodals. Throughout the simulation the mean diameters of particles in the gas and liquid regions, $\langle d \rangle _{\textrm{gas}}$ and $\langle d \rangle_{\textrm{liquid}}$, are recorded. 

We present fractionation data from 4 large simulations with $N = 30000$, where 2 independent initial configurations have been simulated according to both the scalable and non-scalable definitions of the inter-particle potential. In addition, we ran 20 smaller simulations ($N = 4000$), 10 for each potential, to confirm the qualitative nature of the fractionation effects, check for sensitivity to initial configuration or system size, and investigate longer timescales. Finally, having confirmed fractionation in the near-monodisperse regime, we ran 2 simulations at a higher polydispersity $\sigma = 0.2$ and $N = 4000$ in which we expect the strength of fractionation, scaling with $\sigma^2$ \cite{Evans2000}, to increase by an order of magnitude relative to the other simulations. 

\begin{figure}[floatfix]
\includegraphics[width=7.2cm]{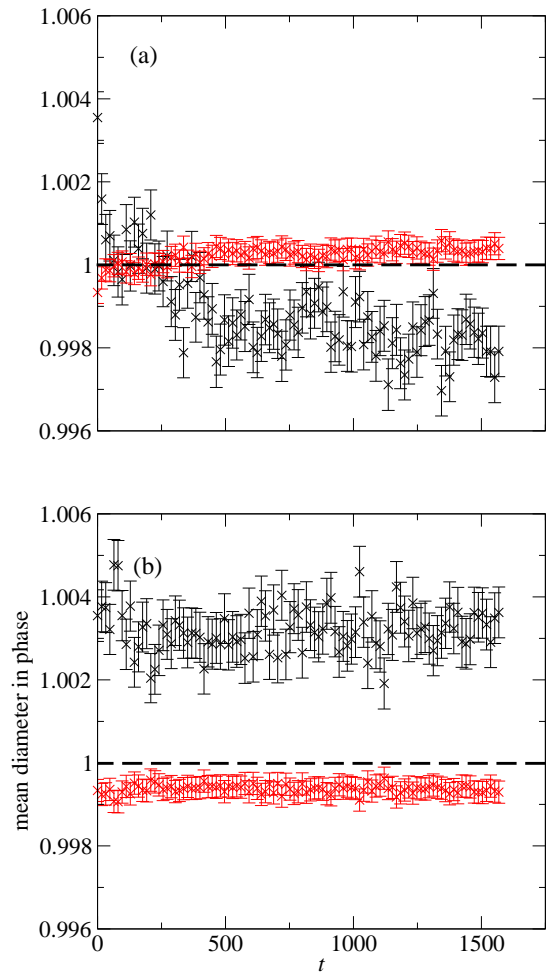}
\caption{\label{fractionation30} (colour online) $\langle d \rangle _{\textrm{gas}}$ (black, large error bars) and $\langle d \rangle _{\textrm{liquid}}$ (grey, red online, small error bars) through time for the two definitions of the inter-particle potential, for 4 simulations with $N = 30000$. The parent mean particle diameter is $\langle d \rangle \equiv 1$. a) For the scalable potential $V_\textrm{{scal}}(r)$ the liquid prefers on average larger particles than the gas. b) The non-scalable potential $V_\textrm{{non-scal}}(r)$ causes the liquid to prefer smaller particles. Error bars represent the standard error on the mean.}
\end{figure}

\begin{figure}[floatfix]
\includegraphics[width=7.2cm]{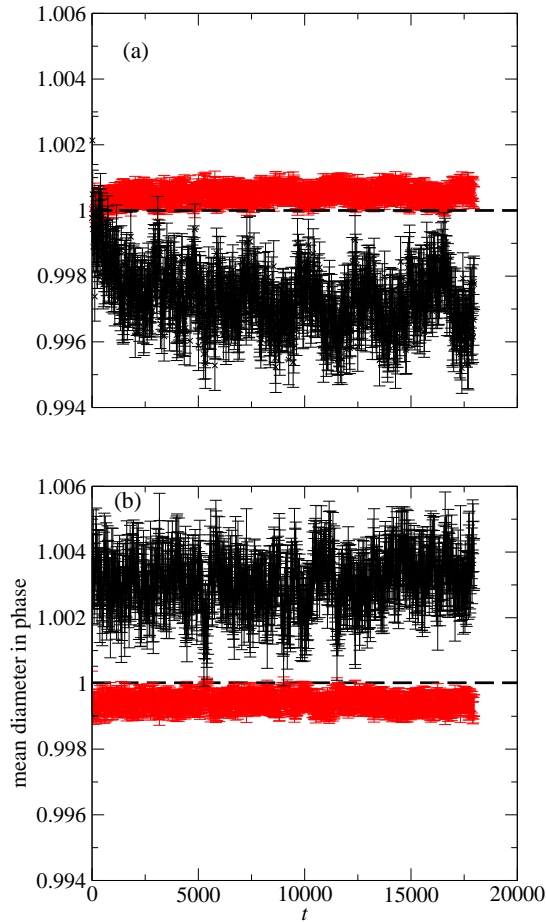}
\caption{\label{fractionation4} (colour online) As FIG.~\ref{fractionation30}, for 20 simulations with $N = 4000$, run for longer times. The fractionation shows no significant change beyond $t \approx 2500$, and is consistent in its magnitude and direction with the data in FIG.~\ref{fractionation30}.}
\end{figure}

\begin{figure}[floatfix]
\includegraphics[width=7.2cm]{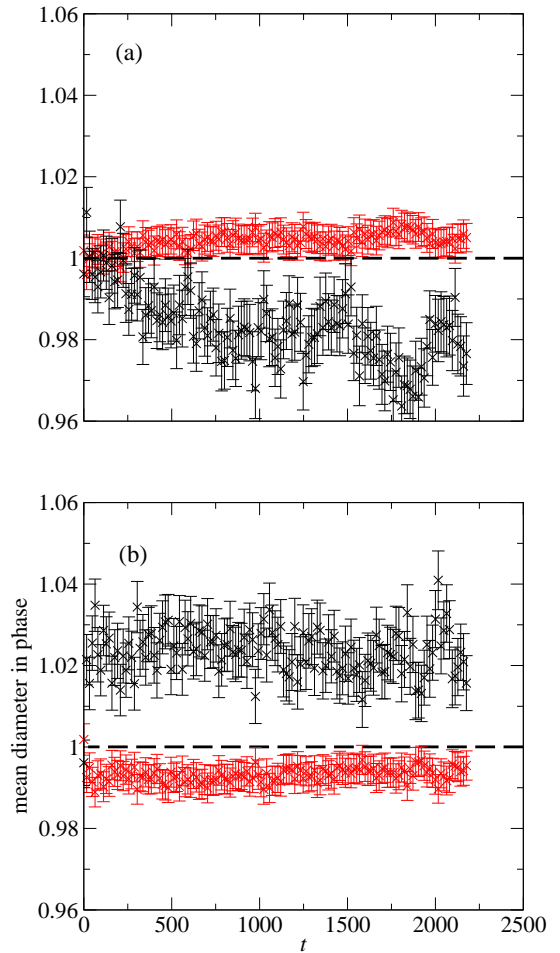}
\caption{\label{fractionationhigh} (colour online) As FIG.~\ref{fractionation30}, for higher polydispersity ($\sigma = 0.2$) simulations. The direction of fractionation is the same as in the lower-$\sigma$ simulations, and the difference in mean size between the phases is increased by approximately an order of magnitude.}
\end{figure}

FIG.~\ref{fractionation30} shows the evolution of $\langle d \rangle _{\textrm{gas}}$ and $\langle d \rangle _{\textrm{liquid}}$ in the large system for the scalable and non-scalable potentials $V_\textrm{{scal}}(r)$ and $V_\textrm{{non-scal}}(r)$. Fractionation of particle sizes is apparent after the first few hundred time units, although for the non-scalable potential the signal is apparently present at very early times (around $50 t_{d}$). We note, however, that at these times the system has yet to properly separate into distinct gas and liquid regions, as shown in FIG.~\ref{snapshots}, so that the signal may be significantly influenced by the presence of many particles whose neighbour counts lie on the threshold between `gas' and `liquid' according to our definition. For such `threshold' particles, a marginally greater size may exclude neighbours from the local region, leading to the particle being counted as gas, whereas at later times, the system has successfully formed distinct phases with neighbour counts very much less and very much higher than the threshold. This measurement effect, if present, seems to be minor for the polydispersities studied here, in that opposite fractionation effects between the two potentials are resolved once the system has formed distinct phases. For this reason, though, we refrain from drawing firm conclusions from the fractionation signal at early simulation times.

As shown in FIG.~\ref{fractionation4}, the fractionation in the smaller, longer-time simulations agrees within error with the data from the larger simulations. Thereafter, the fractionation in these simulations shows no detectable change beyond $t \approx 2500$ up to the maximum time simulated. FIG.~\ref{fractionationhigh} confirms the same qualitative fractionation at higher polydispersity, with the difference in mean size between the phases increased by approximately an order of magnitude as compared to the $\sigma = 0.06$ case. 

Since fractionation relies on self-diffusion of individual particles between the phases, whereas bulk phase separation takes place via faster, collective rearrangement of particle density \cite{Warren1999}, it might be expected that the speed of spinodal decomposition would negate any possibility of fractionation in the early stages of phase separation, if the particles could not rearrange to the required degree faster than the spinodal texture coarsens. Therefore, while our results leave open the question of further `ripening' of the phase composition on much longer timescales, it is significant that fractionation is apparent almost immediately after spinodal decomposition begins. 

Strikingly, there is a qualitative dependence on the choice of inter-particle potential -- the non-scalable potential causes the liquid to accept smaller particles, whereas the scalable potential causes the smaller particles to be incorporated into the gas. Recall that $V_\textrm{{scal}}(r)$ and $V_\textrm{{non-scal}}(r)$ have the same well depth and mean range -- they differ only in the \textit{precise dependence} of their range upon the mean diameter $d_{ij}$ of the particles between which the potential is acting. The two potentials coincide by definition in the monodisperse case (for which they produce exactly identical simulation trajectories), but are `split' by a mild degree of polydispersity so as to produce opposite fractionation effects.

We can explain this surprising dependence on inter-particle potential with an equilibrium theory of polydispersity \cite{Evans2000}. While the system being simulated here is clearly not at equilibrium, it is reasonable to invoke such a theory in order to give some indication of what metastable state the system is relaxing towards. In Ref.~\cite{Evans2000}, perturbation of a monodisperse reference system is used to predict equilibrium properties of polydisperse systems, requiring as input only properties of that monodisperse reference. The approach becomes exact in the limit of weak polydispersity, i.e. if $\langle\epsilon^{2} \rangle$ is small where $\epsilon$ is a particle's deviation from the mean size $\langle d \rangle$, in units of $\langle d \rangle$. The applicability of that theory is the reason for our primarily focusing on the low-$\sigma$ regime in the simulations.

For fractionation, the key thermodynamic potential is $A(\rho) = \rho {d\mu ^{\textrm{ex}} (\epsilon)} / {d\epsilon}$, a function of number density $\rho$ describing how the excess chemical potential for a particle in a monodisperse system varies with $\epsilon$. Once $A(\rho)$ and the polydispersity $\sigma$ of the parent distribution are known, the fractional difference in mean particle size between the phases is given by:

\begin{equation}
[\langle\epsilon\rangle]^{l}_{g} = - [A/\rho ]^{l}_{g} \sigma^{2}
\label{eqn:eqmliquid}
\end{equation}

\noindent where the labels $l$, $g$ indicate quantities evaluated at the metastable coexistence densities of the liquid and gas phases respectively. Thus, if the difference in $A/\rho$ between the two phases, $[A/\rho ]^{l}_{g}$, is negative, the predicted mean size in the metastable liquid phase will be greater than that in the gas. The $\sigma^2$ dependence is consistent with the difference of around one order of magnitude in the fractionation between the $\sigma = 0.2$ and $\sigma = 0.06$ simulations.

To find $A$, an approximate free energy density $f$ is required for the monodisperse square well fluid. We use the attractive part of the interaction potential $U(r) = -u$ as a perturbation to a hard-sphere reference system with the Percus-Yevick radial distribution function $g_{\textrm{HS}}(r)$, giving

\begin{equation}
f \approx  f_{\textrm{HS}} + \frac{\rho^2}{2} \int \! 4\pi r^{2} g_{\textrm{HS}}(r) U(r) \, dr
\label{eqn:hsperturbation}
\end{equation}

\noindent where $f_{\textrm{HS}}$ is the Carnahan-Starling hard-sphere free energy density \cite{Carnahan1969} and the integral is over the square well range.

\begin{figure}[floatfix]
\includegraphics[width=7.2cm]{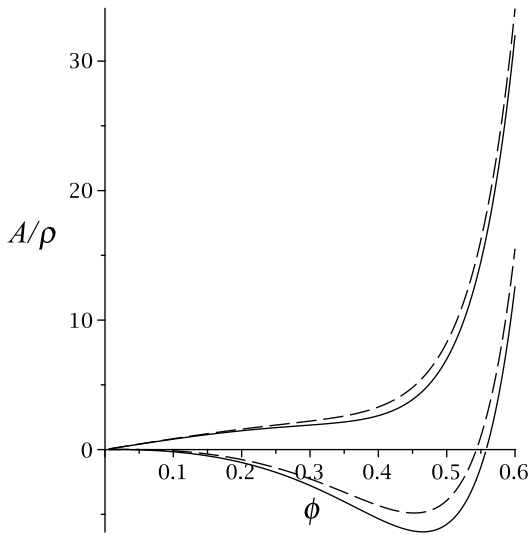}
\caption{\label{Arho} $A/\rho$ shown as a function of volume fraction $\phi$ for a monodisperse reference system with $u = 1.82$ (solid lines), used to predict the fractionation of particle size at gas-liquid coexistence \cite{Evans2000}. Dashed lines show $u = 1.74$ for comparison. The lower two curves are for the scalable interaction potential $V_\textrm{{scal}}(r)$, while the upper two curves correspond to the non-scalable potential $V_\textrm{{non-scal}}(r)$.}
\end{figure}

For  $V_\textrm{{scal}}(r)$, finding $[A/\rho ]^{l}_{g}$ is made simpler because the definition is such that all lengths, including the range, scale with the hard core radius. In this special case, it can be shown \cite{Evans2000} that $A = 3(P - \rho)$ where $P$ is the osmotic pressure calculated from the free energy density. FIG.~\ref{Arho} shows that, for the scalable potential, $A/\rho$ decreases across the gas-liquid coexistence region ($\phi_g \approx 0.05$ to $\phi_l \approx 0.40$) so that $[A/\rho ]^{l}_{g} < 0$, correctly predicting the observed fractionation of larger particles into the liquid phase in FIG.~\ref{fractionation30} a). For $V_\textrm{{non-scal}}(r)$, $A/\rho$ increases monotonically, $[A/\rho ]^{l}_{g} > 0$ and the liquid contains on average \textit{smaller} particles, in qualitative agreement with FIG.~\ref{fractionation30} b). 

This is intuitively reasonable: from Equations \ref{eqn:scalablesquarewell} and \ref{eqn:nonscalablesquarewell}, it is clear that the range of $V_\textrm{{scal}}(r)$ increases more strongly with $d_{ij}$ than that of $V_\textrm{{non-scal}}(r)$. In the non-scalable case, increasing the diameter of particles in the liquid results in a greater increase in chemical potential than in the gas -- the incorporation of larger particles into the liquid is unfavourable. In the scalable case, the marginally stronger dependence of the well range on $d_{ij}$ is such that an increase in particle diameter \textit{reduces} chemical potential, as more particles can fit into the square wells -- this effect is enhanced in the liquid, compared with the gas, due to its smaller inter-particle spacing, so that the liquid now favours larger particles.

For a quantitative comparison, we use Equation~\ref{eqn:eqmliquid} to estimate the extent of `equilibrium' fractionation at gas-liquid coexistence by evaluating $[A/\rho ]^{l}_{g}$ from the  curves in FIG.~\ref{Arho}. Taking the coexistence volume fractions from \cite{Liu2005}, $\phi_g \approx 0.05$ and $\phi_l \approx 0.40$, we find $[A_{\textrm{scal}}/\rho ]^{l}_{g} = -5.31$ and $[A_{\textrm{non-scal}}/\rho ]^{l}_{g} = 2.20$, giving for the $\sigma = 0.06$ system

\begin{equation}
[\langle d \rangle_{\textrm{scal}}]^{l}_{g}  = 0.019 , [\langle d \rangle_{\textrm{non-scal}}]^{l}_{g}  = - 0.0079~. 
\label{eqn:liqequil}
\end{equation}

\noindent It appears that, while the fractionation observed on the simulated timescale is in the \textit{direction} of equilibrium, it is substantially weaker than that required to minimise the free energy of the metastably coexisting phases, consistent with the hypothesis that the compositions of the phases would typically relax far more slowly than their overall densities \cite{Warren1999}. Given this, it is all the more remarkable that the qualitative sensitivity of fractionation to the distinction between $V_\textrm{scal}(r)$ and $V_\textrm{non-scal}(r)$ is present even at the very early stages of phase separation. 

\section{\label{sec:conclusion}Conclusion}

Whilst dynamical simulations of crystallisation in bidisperse mixtures have previously found evidence of fractionation \cite{Williams2008} these are, to our knowledge, the first such measurements in the early-stage phase separation of a truly polydisperse model colloidal fluid. The equilibrium theory used here has been tested in experimental scenarios \cite{Evans1998, Erne2005, Fairhurst2004} in which fractionation measurements are typically taken after a long period of equilibration. For a typical colloidal particle of radius $\approx 100$~nm in water, the diffusion time is $t_{d} \sim 10^{-3}$ s in which case the simulation time used here represents the first few seconds of phase separation. Experimentally, measurements in these early stages are difficult or impossible in the absence of large single-phase regions or when the system is evolving quickly. 

The data presented show that the beginnings of fractionation can take effect from the outset of spinodal decomposition, but with a smaller magnitude than predicted by equilibrium calculations. It must be noted that those calculations used a rather approximate perturbative expression for the free energy, in order to simplify the subsequent differentiation, and that their precise results may be subject to quantitative change if a more sophisticated expression were used. Nevertheless, the theory used was appropriate for our purposes, correctly predicting the qualitative dependence on inter-particle potential that is observed in the simulations.

As discussed in Section~\ref{sec:simulation}, Hydrodynamic Interactions (HI) are important for spinodal decomposition in real systems, but were neglected in our simulations in order to reduce computational expense. We argued that, especially since HI may be screened to a greater or lesser extent, the question of how fractionation takes place in their absence is a pertinent one. Further work will be required to incorporate HI into simulations still large enough to gather the required statistics for fractionation. 

Although we focused on the low-polydispersity regime in which the perturbative theory used here becomes exact, simulations at higher polydispersity confirmed the same qualitative physics, accompanied by an order of magnitude increase in the difference in mean particle size between phases, as expected from Equation~\ref{eqn:eqmliquid}. We have also found an example of the subtle interplay between polydispersity and other system parameters, in that the choice between two inter-particle potentials which are identical in the monodisperse case can qualitatively change the predicted and observed direction of fractionation in a slightly polydisperse system. 

The latter finding starkly illustrates the care that must be taken as the treatment of polydisperse systems becomes more sophisticated: under a `near-monodisperse' approximation, the choice between the two inter-particle potentials studied here might have been expected to have a negligible effect on the results and therefore be made simply to maximise theoretical expediency. On the contrary, we have shown that this choice can dramatically affect (and in this case, reverse) the fractionation observed, a fact that must be taken into account if precise knowledge and control of the composition and dependent properties of the two phases is desired. Further, the fact that dynamical simulation can measure fractionation at the early stages of phase separation highlights the detailed role it may play in bridging the gap between equilibrium theories of colloidal polydispersity and the behaviour of real, dynamical systems.

\begin{acknowledgments} 
The authors thank W.~C.~K.~Poon, N.~B.~ Wilding and S.~Liddle for illuminating discussions. JJW was supported by an EPSRC DTG award, and acknowledges the assistance of D.~M.~Whitley in accessing the computing resources used to carry out the work.
\end{acknowledgments}

\bibliography{Main}

\end{document}